\documentclass[twocolumn]{aastex631}
\usepackage{CJK}
\usepackage{multirow}

\newcommand{\bdv}[1]{\mbox{\boldmath$#1$}}
\newcommand{\mps}{{\rm m\,s^{-1}}}

\shorttitle{Planet Intrinsic Multiplicities}
\shortauthors{Wei Zhu}
\graphicspath{{./}{figures/}}

\begin{document}
\begin{CJK*}{UTF8}{gbsn}

\title{The intrinsic multiplicity distribution of exoplanets revealed from the radial velocity method}

\author[0000-0003-4027-4711]{Wei Zhu (祝伟)}
\affiliation{Department of Astronomy, Tsinghua University, Beijing 100084, China} \email{weizhu@tsinghua.edu.cn}



\begin{abstract}
Planet multiplicities are useful in constraining the formation and evolution of planetary systems but usually difficult to constrain observationally. Here, we develop a general method that can properly take into account the survey incompleteness and recover the intrinsic planet multiplicity distribution. We then apply it to the radial velocity (RV) planet sample from the California Legacy Survey (CLS). Within the 1\,au (10\,au) region, we find $21 \pm 4\%$ ($19.2 \pm 2.8\%$) of Sun-like stars host planets with masses above $10\,M_\oplus$ ($0.3\,M_{\rm J}$), about 30\% (40\%) of which are multi-planet systems; in terms of the RV semi-amplitude $K$, $33 \pm 7\%$ ($25 \pm 3\%$) of Sun-like stars contain planets of $K>1\,\mps$ ($3\,\mps$), and each system hosts on average $1.8 \pm 0.4$ ($1.63 \pm 0.16$) planets. We note that the hot Jupiter rate in the CLS Sun-like sample is higher than the consensus value of $\sim$1\% by a factor of about three. We also confirm previous studies on the correlation between inner ($<1\,$au) and outer ($>1\,$au) planets.
\end{abstract}



\section{Introduction} \label{sec:intro}

Multi-planet systems are common. Of the thousands of exoplanets (or planet candidates) found by the \emph{Kepler} mission \citep{Borucki:2010}, about half of them are known to have siblings \citep[e.g.,][]{Thompson:2018, Berger:2018}. This fraction further increases dramatically after the transit probability is properly taken into account \citep[e.g.,][]{Lissauer:2011, Tremaine:2012, Fabrycky:2014, Zhu:2018, He:2020}. Besides the planet sample from transit surveys, multi-planet systems have also been found to be common via (combinations of) other detection techniques \citep[e.g.,][]{Wright:2009, Gould:2010, Knutson:2014, Bryan:2016, ZhuWu:2018}.

Planet multiplicity conveys rich information about the system formation and evolution. To the zeroth order, planets around the same host are formed out of the same disk, and thus whether and how different types of planets can reside in the same system can provide important constraints on their formation channels and/or locations \citep[e.g.,][]{Knutson:2014, Huang:2016, ZhuWu:2018, Bryan:2019}. Additionally, the dynamical interactions among planets may have also imprinted signatures on the present-day architecture of the planetary system \citep[e.g.,][]{Chatterjee:2008, Juric:2008, PuWu:2015}. We refer interested readers to \citet{Dawson:2018} and \citet{ZhuDong:2021} for some recent reviews on the related topics.

To better connect observations with theoretical models of planet formation and evolution, we need to go beyond the commonly used planet occurrence rates and constrain directly the intrinsic multiplicity distribution of planetary systems, that is, the fractions of stars with different numbers of planets within the given parameter space. This requires additional modeling effort to properly take into account the observational biases, as they may affect different multiplicities at different levels.

Because the \emph{Kepler} survey provided so far the largest uniform planet sample, previous studies have relied on the \emph{Kepler} transit sample to deriving the intrinsic multiplicity distribution \citep[e.g.,][]{Lissauer:2011, FangMargot:2012, Zhu:2018, Mulders:2018, Zink:2019, He:2020}. However, the transit method is compromised by the strong degeneracy between the intrinsic multiplicity and the mutual inclination distribution, which cannot be resolved without external constraints (e.g., \citealt{Lissauer:2011, Tremaine:2012, Zhu:2018}; see \citealt{ZhuDong:2021} for further discussions on this issue).

In this work, we make use of the recently released planet sample from the radial velocity (RV) survey---the California Legacy Survey (CLS, \citealt{Rosenthal:2021})---and derive the intrinsic multiplicity distribution of planets around Sun-like stars. 
We describe the general method to model the multiplicity distribution in any RV survey in Section~\ref{sec:method}. We then apply this method to the CLS sample and report the results in Section~\ref{sec:cls}. A discussion about the method as well as the results from the CLS sample is given in Section~\ref{sec:discussion}.

\section{Method} \label{sec:method}

\subsection{Notations}

We follow the general framework of \citet{Tremaine:2012} to derive the intrinsic multiplicity distribution from an exoplanet survey with limited sensitivities. We use $\bdv{F}$ for the intrinsic multiplicity vector, with each component $F_k$ representing the fraction of stars with $k$ planets in the target parameter space.
\footnote{Throughout this paper the letter $k$ is used for the number of intrinsic planets, regardless of whether they are detected or not, whereas the letter $j$ is used for the number of planets that are actually detected.}
The observed multiplicity vector is $\bdv{N}$, with each component $N_j$ representing the number of stars with $j$ detected planets. With a statistical sample of $N_\star$ stars searched in the survey, a relation can be established
\begin{equation} \label{eqn:nsf}
    \bdv{N} = N_\star \bar{\mathcal{S}} \bdv{F} ,
\end{equation}
or equivalently
\begin{equation}
     N_j = N_\star \sum_{k=0} \bar{S}_{jk} F_k.
\end{equation}
Here $\bar{\mathcal{S}}$ is a matrix that captures the survey detection sensitivity, averaged over all stars in the sample. Each entry of the matrix, $\bar{S}_{jk}$, quantifies the (averaged) probability that the survey identifies $j$ planet detections in a typical system of $k$ planets \citep{Tremaine:2012}.

Under the general assumption that the detections of individual planets around the same host are independent, the sensitivity matrix $\mathcal{S}$ of the $l$-th star in the sample is given as
\begin{equation} \label{eqn:s_matrix}
    S_{jk} (p_l) = \left\{
    \begin{array}{ll}
        \frac{k!}{j! (k-j)!} p_l^j (1-p_l)^{k-j} , &  j \leq k \\
        0 , & j > k
    \end{array} \right. ,
\end{equation}
where $p_l$ is the detection probability of the $l$-th star to a typical planet in the pre-defined parameter space $\Delta \bdv{\theta}$
\begin{equation} \label{eqn:p_definition}
    p_l \equiv \frac{\int_{\Delta \bdv{\theta}} f(\bdv{\theta}) s_l (\bdv{\theta}) d\bdv{\theta}}{\int_{\Delta \bdv{\theta}} f(\bdv{\theta}) d\bdv{\theta}} .
\end{equation}
Here $\bdv{\theta}$ denotes a set of parameters that define the properties of the planet.
Note that the evaluation of $p_l$ takes into account not only the detection sensitivity of the star, $s_l(\bdv{\theta})$, but also the intrinsic planet distribution of all stars, $f(\bdv{\theta})$. The mean sensitivity of the survey is thus
\begin{equation}
    \bar{\mathcal{S}} = \frac{1}{N_\star} \sum_{l=1}^{N_\star} \mathcal{S}(p_l) .
\end{equation}

The frequency of planets, or the average number of planets per star, is
\begin{equation}
    \bar{n}_{\rm p} \equiv \sum_{k=1} k F_k .
\end{equation}
With the above equations it is straightforward to show that
\begin{equation} \label{eqn:np}
    \bar{n}_{\rm p} = \frac{1}{N_\star \bar{p} } \sum_{j=1} j N_j ,
\end{equation}
where the averaged detection probability
\begin{equation}
    \bar{p} \equiv \frac{1}{N_\star} \sum_{l=1}^{N_\star} p_l .
\end{equation}
That is, the derivation of the frequency (or occurrence rate) of planets is independent of the multiplicity distribution. This is consistent with the original assumption that the detections of individual planets are independent. Additionally, as the planet frequency only depends on the averaged detection probability $\bar{p}$, it provides a practical way to derive the intrinsic planet distribution $f(\bdv{\theta})$. This last quantity is central to the calculation of the per-star detection probability $p_l$ (Equation~\ref{eqn:p_definition}). We will further demonstrate it in Section~\ref{sec:cls}.

Another quantity of interest to the exoplanet statistics is the frequency of planetary systems, or the fraction of stars with (at least one) planets,
\begin{equation}
    F_{\rm p} \equiv \sum_{k=1} F_k = 1-F_0 .
\end{equation}
Unlike the derivation of $\bar{n}_{\rm p}$, the derivation of $F_{\rm p}$ depends on planet multiplicity.

The ratio between the frequency of planets and the frequency of planetary systems gives the average number of planets per planetary system, or average multiplicity \citep{ZhuDong:2021}
\begin{equation}
    \bar{m}_{\rm p} \equiv \frac{\bar{n}_{\rm p}}{F_{\rm p}} .
\end{equation}

\subsection{Maximum likelihood analysis} \label{sec:modeling}

With a given intrinsic multiplicity vector, $\bdv{F}$, and the sensitivity matrix, $\bar{\mathcal{S}}$, one can compute the expected multiplicity distribution, $\bar{\bdv{N}} \equiv \{\bar{N}_j\}$, based on Equation~(\ref{eqn:nsf}). The detections of individual systems are best described as Poisson processes, and thus the likelihood of having the observed multiplicity distribution, $\bdv{N} \equiv \{N_j\}$, is given by
\begin{equation} \label{eqn:likelihood}
    \mathcal{L} = \prod_{j=0} \frac{\bar{N}_j^{N_j} \exp(-\bar{N}_j)}{N_j!} ,
\end{equation}
and its logarithm
\begin{equation}
    \ln{\mathcal{L}} = \sum_{j=0} \left( N_j\ln{\bar{N}_j} - \bar{N}_j \right)  + {\rm constant} .
\end{equation}
The constant term is ignored hereafter. The above likelihood is maximized when $\bar{\bdv{N}}=\bdv{N}$, leading to the solution
\begin{equation} \label{eqn:solution_analytic}
    \bdv{F} = \frac{1}{N_\star} \bar{\mathcal{S}}^{-1} \bdv{N} .
\end{equation}
The above analytic solution does not take into account the physical constraints on the intrinsic multiplicities, namely $0 \leq F_k \leq 1$. Nevertheless, it provides a simply and useful way to check the numerical results and investigate the impact of input parameters on the derived frequencies.

One can also derive/constrain the intrinsic multiplicities in a numerical way. This allows to impose the physical constraints and derive the uncertainties of the intrinsic multiplicities. The data to be fitted by the model is the observed multiplicity distribution, $(N_0, N_1, N_2, \dots)$, and the free parameters of the model are $(F_1, F_2, \dots)$. Note that the first component of $\bdv{F}$, $F_0$, is not a free parameter of the model, because it can be derived by $F_0 = 1 - \sum_{j=1} F_j$. Therefore, the number of observational constraints is always larger than the number of free parameters, and thus the model can be constrained.

The observed multiplicity distribution can in principle be extended to infinity, which cannot be handled in practice, so where to truncate it? Theoretically, mass conservation of the planetary material and long-term dynamical stability arguments do not allow a system to host infinite number of planets. However, such arguments do not give deterministic criterion on the maximum number of planets per system, so we rely on the statistical argument and regard it as an issue of model selection. The likelihood of the model is optimized under Equation~(\ref{eqn:solution_analytic}), and the inclusion of higher multiplicities with zero detections no longer enhances it. Therefore, both intrinsic and observed multiplicity distributions are truncated at the highest multiplicities with non-zero detections.
\footnote{A few criteria, such as the Bayesian Information Criterion (BIC), have been proposed to quantify the superiority of one model over another. However, the commonly used form of BIC assumes that the number of data points substantially exceeds the number of model parameters. This is not true in the present case.}
We will discuss the robustness of our results against the choice of the maximum multiplicity in the modeling, $k_{\rm max}$, in Section~\ref{sec:kmax}.

We derive the uncertainties on the intrinsic multiplicities in the following way. We generate a fine grid around the maximum-likelihood solution in Equation~(\ref{eqn:solution_analytic}) and calculate the likelihood value at each grid point. The $n$-$\sigma$ confidence interval is then defined by grid points whose likelihoods are no more than $\Delta\ln{\mathcal{L}} = n^2/2$ from the maximum value.

\section{Application to the CLS sample} \label{sec:cls}

\subsection{Sample description} \label{sec:sample}

\begin{figure}
    \centering
    \includegraphics[width=1.0\columnwidth]{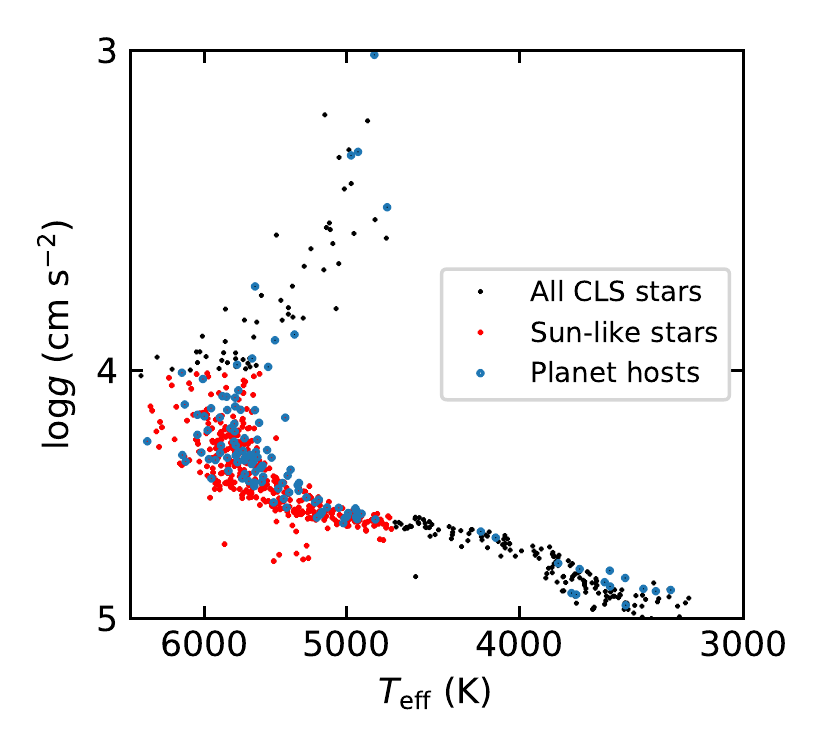}
    \caption{Surface gravity ($\log{g}$) vs.\ effective temperature ($T_{\rm eff}$) for stars in the CLS sample. Here black dots are all CLS stars, red dots are the Sun-like stars according to the selection criteria of Section~\ref{sec:sample}, and the blue circles mark the CLS stars with planet detections.}
    \label{fig:hr}
\end{figure}

\begin{figure*}
    \centering
    \includegraphics[width=0.7\textwidth]{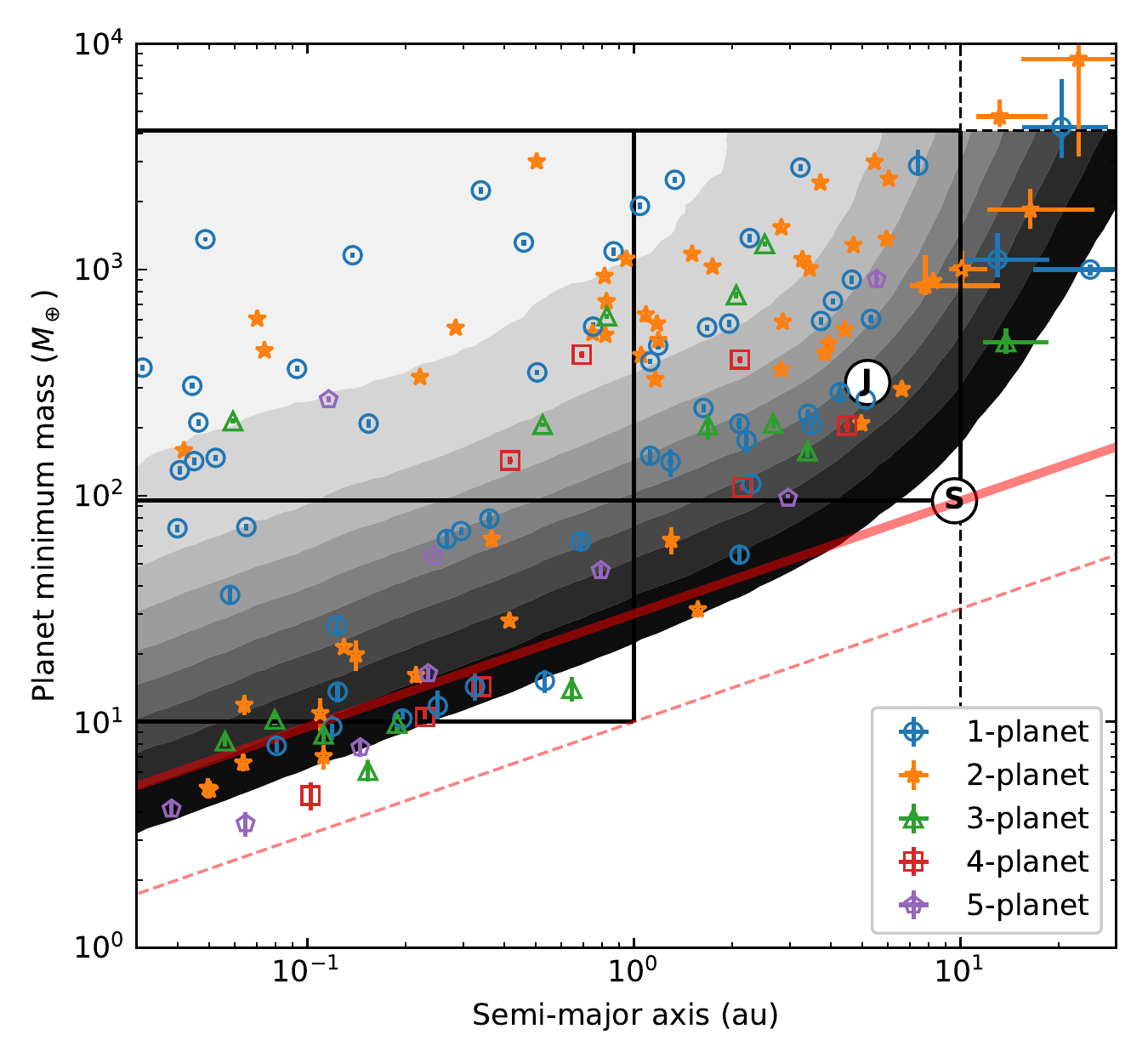}
    \caption{The distribution of CLS planets with Sun-like hosts in the minimum mass vs.\ semi-major axis plane. Planets with different observed multiplicities are differentiated with different labels and colors. The contours are the survey completeness curves of 10\% (darkest) to 90\% (lightest), constructed based on the injection--recovery simulations of all Sun-like stars in the CLS sample \citep{Rosenthal:2021}. The vertical dashed line indicates the outer boundary (10\,au) that is considered in this work, and the horizontal dashed line indicates the upper mass limit ($13\,M_{\rm J}$). The positions of Jupiter and Saturn are also shown. The regions of three types of planets defined in Section~\ref{sec:cls} are given as the black boxes. The red solid and dashed lines denote the RV semi-amplitudes of 3 and $1\,\mps$, respectively. Note that one additional planet, 55 Cnc e, with minimum mass $m\sin{i}=9.4\,M_\oplus$ and semi-major axis $a=0.02\,$au, is slightly beyond the limit and thus does not appear on the plot. }
    \label{fig:planets}
\end{figure*}

\begin{figure*}
\centering
\includegraphics[width=0.95\textwidth]{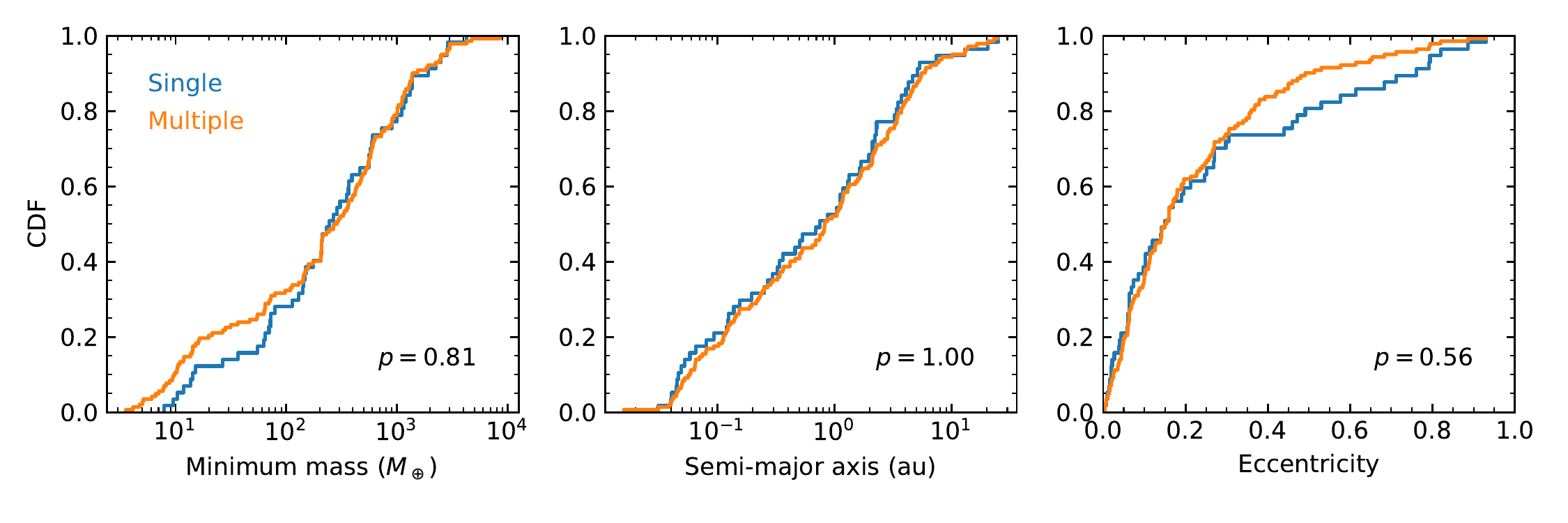}
\caption{The cumulative distribution functions (CDFs) of planetary parameters for planets from systems with one planet detection (blue curves) and more than one planet detections (orange curves). In the left, middle, and right panels are shown the minimum mass, the semi-major axis, and the orbital eccentricity, respectively. The $p$ value of the two-sample KS test is indicated in each of these comparisons.}
\label{fig:comparison}
\end{figure*}

The CLS sample contains 178 RV planet detections around 719 stars \citep{Rosenthal:2021,Fulton:2021}. The CLS stars cover a broad range in stellar properties, including FGKM-type dwarf and giant stars. For this study, we will focus on the Sun-like stars, which are defined as stars with effective temperature, $T_{\rm eff}$, in the range of 4700--6500\,K and surface gravity $\log{g}>4.0$. Applying these cuts to the CLS sample, we are left with 474 Sun-like stars. These include 92 known planet hosts, and the total number of CLS planet detections is 142. Figure~\ref{fig:hr} shows the selected Sun-like stars and planet hosts in comparison to the overall CLS sample.

We reconstruct the CLS completeness curves for the selected stars based on the per-star injection--recovery simulations of \citet{Rosenthal:2021}. The averaged completeness curves are shown in Figure~\ref{fig:planets} together with the CLS planet detections around Sun-like hosts. Our sample has $(382, 57, 26, 5, 2, 2)$ systems with $(0, 1, 2, 3, 4, 5)$ planet detections.

In principle, one can use the overall observed multiplicity distribution and the estimated completeness curves to constrain the intrinsic multiplicity vector of planets across the whole mass--semi-major axis parameter space. However, given the limited sensitivity of CLS, such an attempt is determined to be very sensitive to assumptions about the unknown distributions of low-mass, long-period planets. We therefore limit ourselves to certain regions of the parameter space. Specifically, we identify the following types of planets (see Figure~\ref{fig:planets} for an illustration):
\begin{itemize}
    \item \emph{Close-in giant planets}: planets with semi-major axis $a<1\,$AU and minimum mass $m\sin{i}$ in the range 0.3--13$\,M_{\rm J}$;
    \item \emph{Close-in intermediate-mass planets}: planets with semi-major axis $a<1\,$AU and minimum mass $m\sin{i}$ in the range 10$\,M_\oplus$--0.3$\,M_{\rm J}$;
    \item \emph{Cold giant planets}: planets with semi-major axis $a$ in the range 1--10\,AU and minimum mass $m\sin{i}$ in the range 0.3--13$\,M_{\rm J}$;
\end{itemize}
We also study the combinations of the above planet types, namely \emph{close-in giant and intermediate-mass planets} and \emph{all giant planets} (including close-in and cold giant planets). Additionally, for the purpose of future RV observations, we also study the planet distribution in the RV semi-amplitude $K$ vs.\ semi-major axis plane and derive the multiplicity distributions for planets with $K>3\,\mps$ and $a<10\,$au as well as with $K>1\,\mps$ and $a<1\,$au. The observed multiplicity distributions for different types of planets are summarized in Table~\ref{tab:multiplicity}. 

The method outlined in Section~\ref{sec:method} to derive the intrinsic planet multiplicity has assumed that the properties (e.g., orbital periods, mass, etc) and detections of individual planets around the same host are mutually independent. Given data with high enough quality, the detections of multiple planets in the same system are indeed independent. In practice, however, the detectability of a new planet is dependent on the properties of the detected planets. One well-known example is the degeneracy between one eccentric planet and two near-circular planets in 2:1 resonant orbits \citep[e.g.,][]{AngladaEscude:2010, Wittenmyer:2019}. The recent analysis by \citet{Boisvert:2018} found that about 25\% of published RV systems might suffer from this degeneracy. This fraction may be taken as an upper limit on the impact of the degeneracy in the CLS sample, given that CLS on average has more data points and covers longer durations.
For systems that may actually suffer from this degeneracy, the hidden planets are usually not massive enough,
\footnote{Specifically, the hidden planet is lighter than the observed one by a factor of $\sim e/2^{1/3}$, where $e$ is the orbital eccentricity of the reported planet.}
and thus our main conclusion about the giant planet multiplicity will not be affected. Future studies and observations will be needed to understand the impact of the degeneracy on the derived intrinsic multiplicity of low-mass planets.

One indirect way to check the validity of the assumption about independence is to compare the parameters of planets from systems with only one planet detection and at least two planet detections. Such a comparison is shown in Figure~\ref{fig:comparison}. The two-sample Kolmogorov-Smirnov (KS) tests yield $p$-values on the order of unity, suggesting that the two distributions under comparison can be considered to be drawn from the same underlying distribution. In other words, the planetary parameters are statistically similar regardless of the number of planet detections. Therefore, it is reasonable to assume that the planet properties and detections are largely independent in multi-planet systems. See also \citet{Tremaine:2012} for further discussion on this separability issue.

Detected planets in the \emph{Kepler} multi-planet systems appear to have similar properties \citep[e.g.,][]{Lissauer:2011, Ciardi:2013}. Whether or not this is of astrophysical origin is still under debate (e.g., \citealt{Millholland:2017, Weiss:2018, He:2019, Zhu:2020, Murchikova:2020}), and we refer to \citet{ZhuDong:2021} for further discussions about this issue. Regardless of the origin, it has marginal impact on the method and results of the current work for two reasons.
First, even if the observed intra-system similarity in \emph{Kepler} systems is astrophysical, such a correlation will be much weaker in typical RV systems, which have on average more massive and fewer planets.
Second, as explained earlier in the paper, we study the multiplicity distribution of specific planet types whose physical properties are already restricted to a relatively narrow range (e.g., all giant planets have very similar sizes). In other words, the hypothetical planets in the hypothetical multi-planet systems implied in our method already have similar properties, even though they are generated in an independent way.
The primary parameter that is affected by assumptions like the intra-system similarity is the detection probability $p$. We will discuss in Section~\ref{sec:p-impact} how our results will be changed with different average detection probabilities.

\subsection{Frequencies of planets} \label{sec:planet-frequency}

\begin{figure*}
    \centering
    \includegraphics[width=0.7\textwidth]{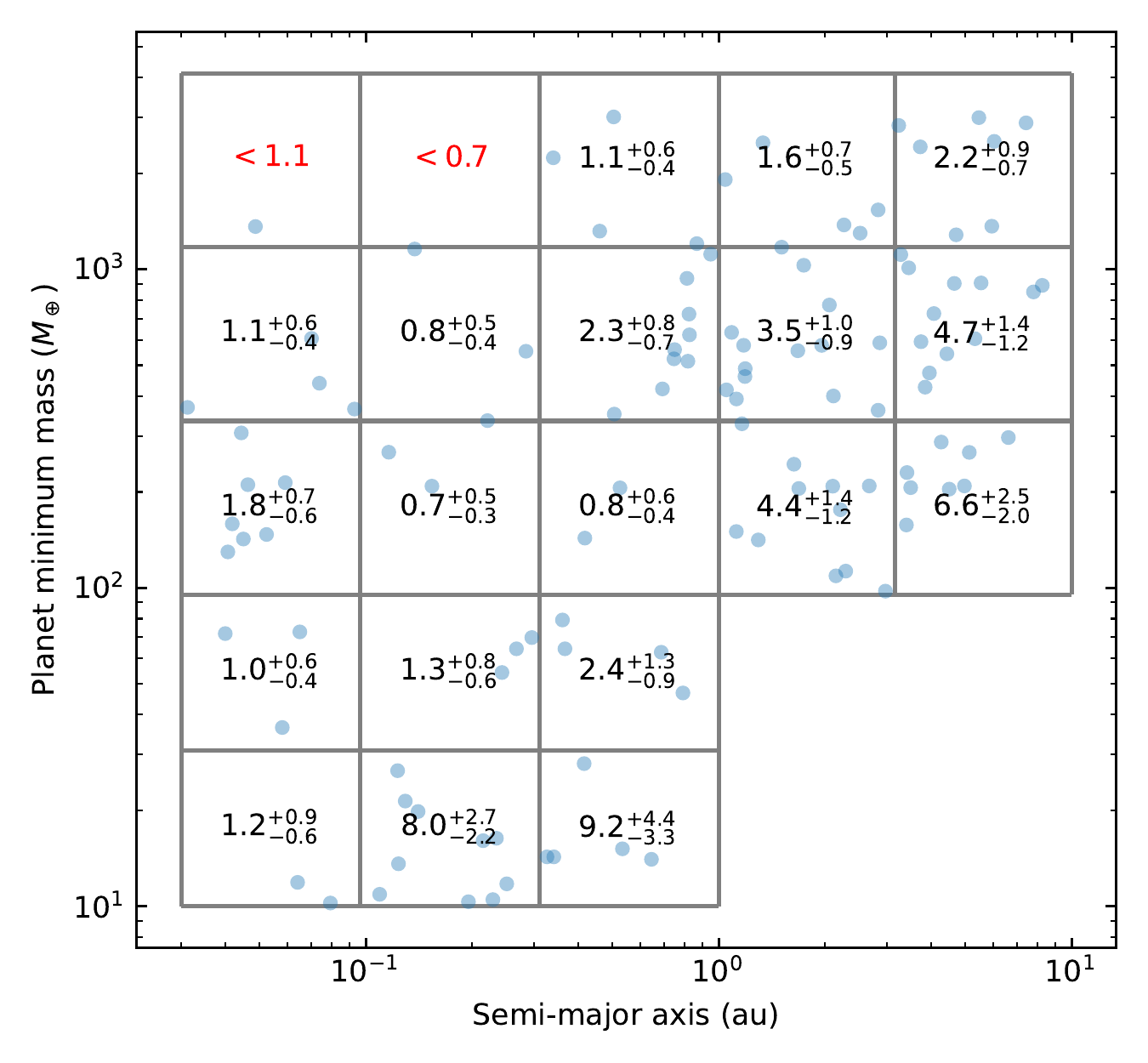}
    \caption{This figure shows in blue dots the planet detections from the CLS Sun-like sample whose properties are within our chosen parameter space. The values within each grid point gives the average number of such planets per 100 Sun-like stars. Error bars gives the 68\% confidence interval, whereas the values in red indicate the 95\% upper limit. The integrated planet frequencies are $\bar{n}_{\rm p}=0.076\pm0.013$, $0.21\pm0.05$, $0.22\pm0.03$ for our definition of close-in giants, close-in intermediate-mass, and cold giant planets, respectively.}
    \label{fig:grid_map}
\end{figure*}

\begin{figure*}
    \centering
    \includegraphics[width=0.7\textwidth]{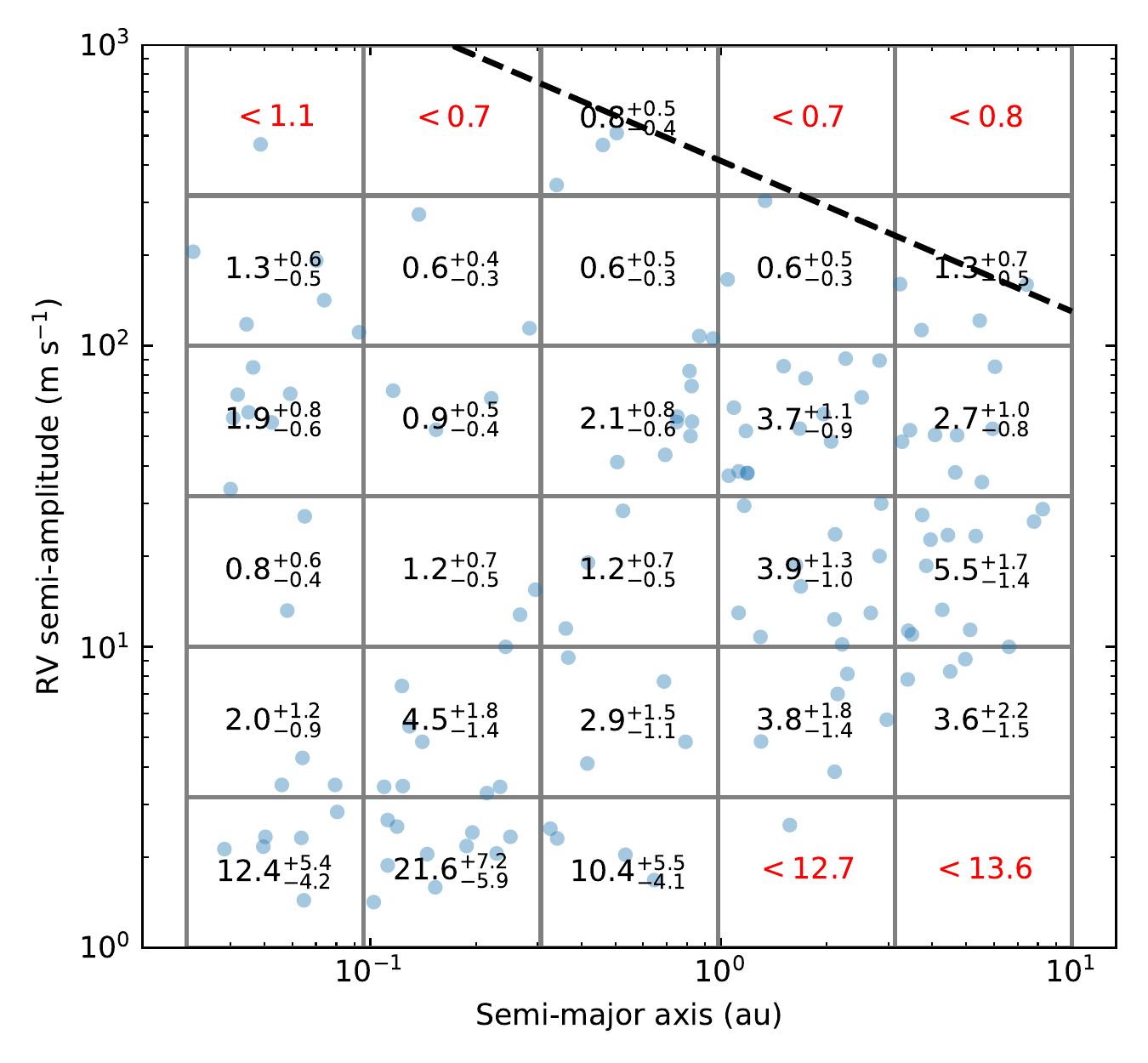}
    \caption{Similar to Figure~\ref{fig:grid_map} except that now the $y$-axis is the RV semi-amplitude $K$. The dashed line indicates the upper mass limit of a planet ($13\,M_{\rm J}$) around a $1\,M_\odot$ host. The frequency of planets with $K>3\,\mps$ and $a<10\,$au is $0.42 \pm 0.04$. The frequency of planets with $K>1\,\mps$ and $a<1\,$au is $0.58\pm0.09$.}
    \label{fig:grid_map_k}
\end{figure*}

\begin{figure}
    \centering
    \includegraphics[width=\columnwidth]{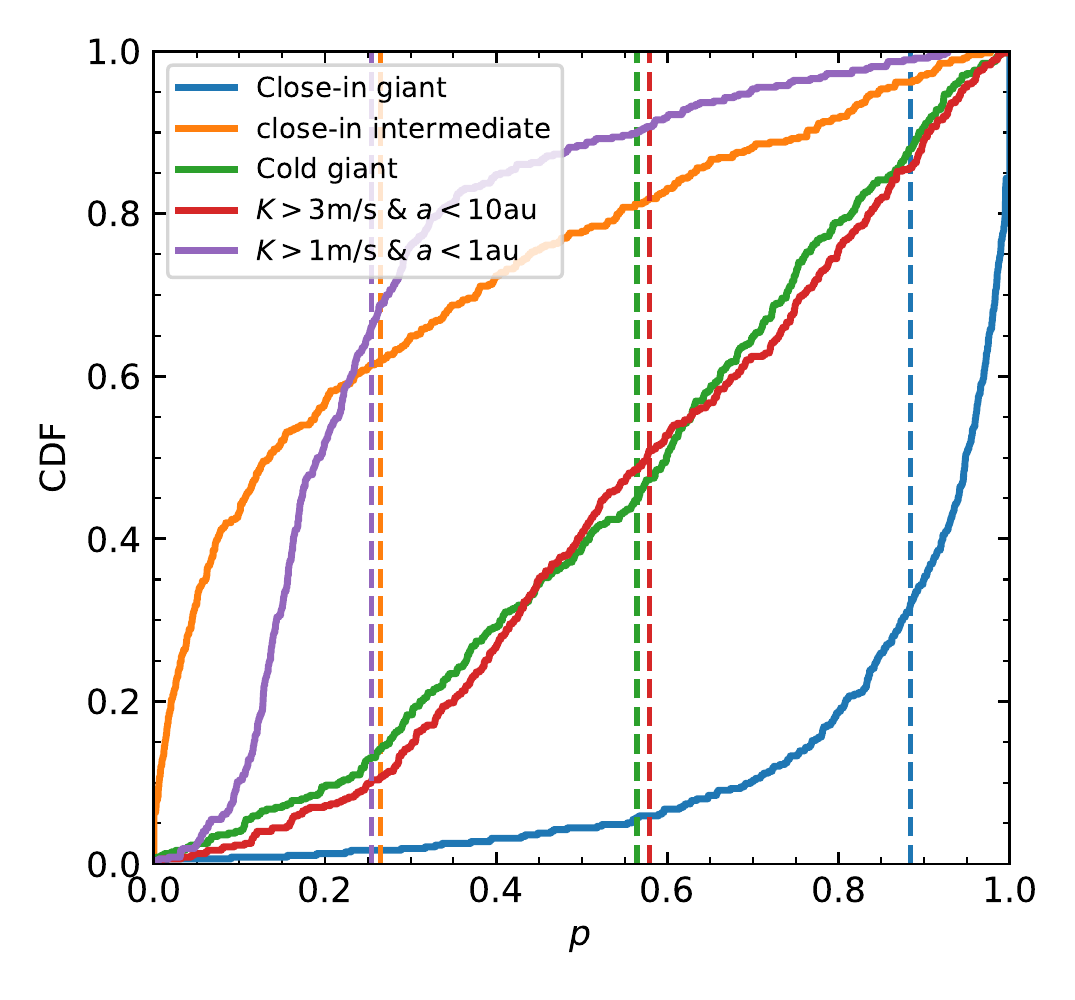}
    \caption{Cumulative distributions of the per-star detection probabilities $p$ for the five chosen planet types. For any given planet type, the vertical dashed line of the same color indicate the average detection probability $\bar{p}$.}
    \label{fig:p-cdfs}
\end{figure}

We first derive the planet frequency, $\bar{n}_{\rm p}$, over the chosen parameter space, in order to determine the mean detection probability $\bar{p}$ as well as the detection probabilities $\{p_l\}$ for individual stars. 

We follow the general approach of \citet{ZhuDong:2021} to derive the planet frequencies. The parameter space is first divided into a combination of grids, with each grid covering an area that is small compared to the overall parameter space but large compared to the positional uncertainties of enclosed planets. Within each grid, the planet distribution function can be treated as a constant, and the average detection probability is given by the mean efficiency of the survey detection. The posterior distribution of the planet frequency within each grid is given by a Gamma distribution. For any grid with at least two planet detections, a log-flat prior on the planet frequency is adopted to determine the median and the 68\% confidence interval. Otherwise, the 95\% upper limit is derived under a linear prior. The results are illustrated in Figures~\ref{fig:grid_map} and \ref{fig:grid_map_k} for planets in the mass vs.\ semi-major axis plane and the RV semi-amplitude vs.\ semi-major axis plane, respectively.

For our definition of close-in giant, close-in intermediate-mass, and cold giant planets, the planet frequency, $\bar{n}_{\rm p}$, is determined to be $0.076\pm0.013$, $0.21\pm0.05$, and $0.22\pm0.03$, respectively. In the $K$ vs.\ $a$ plane, the planet frequency is $0.42\pm0.04$ for planets with $K>3\,\mps$ and $a<10\,$au and $0.58\pm0.09$ for planets with $K>1\,\mps$ and $a<1\,$au. The average detection probabilities for these planet classes can be derived according to Equation~(\ref{eqn:np}). Similarly, the average detection probability for planets defined in the joint parameter space (i.e., close-in CLS planets and all giant planets) as well as those in the RV semi-amplitude vs.\ semi-major axis plane can also be derived. The median values of $p$ for all planet classes are included in Table~\ref{tab:multiplicity}.

In order to derive of the intrinsic multiplicity distribution, we need to calculate the per-star detection probability, $p_l$, which is the detection probability of a typical planet around the $l$-th star in the sample. Making use of the injection--recovery simulations, this is given by
\begin{equation}
    p_l = \frac{\sum_{j'} f(\bdv{\theta}_{j'}) s_l(\bdv{\theta}_{j'})} {\sum_{j'} f(\bdv{\theta}_{j'})} .
\end{equation}
Here $f(\bdv{\theta})$ is the rate map in the specified parameter space (i.e., Figure~\ref{fig:grid_map} for $m\sin{i}$ vs.\ $a$ and Figure~\ref{fig:grid_map_k} for $K$ vs.\ $a$), 
and $s_l(\bdv{\theta}_{j'})$ evaluates whether or not the $j'$-th injected planet around the $l$-th star is detected in the simulation. The summations are performed over all simulated planets that are within the pre-defined parameter space. We show in Figure~\ref{fig:p-cdfs} the distributions of the per-star detection probabilities for the five planet types considered here. As it indicates, this detection probability parameter varies substantially among all stars in the sample, primarily due to the different numbers and quality levels of RV observations for individual stars.

These per-star detection probabilities enable the calculation of the mean sensitivity matrix, $\bar{\mathcal{S}}$, a key ingredient in the derivation of the intrinsic multiplicity distribution.

\subsection{Intrinsic multiplicity distributions}

\begin{deluxetable*}{cccccccc}
\tablecaption{The observed multiplicity distribution and the estimated mean detection probability $p$ for different types of planets in the CLS Sun-like star sample.
\label{tab:multiplicity}}
\tablehead{
\colhead{} & \colhead{\multirow{2}{*}{Close-in giant}} & \colhead{Close-in} & \colhead{\multirow{2}{*}{Cold giant}} & \colhead{\multirow{2}{*}{All giant}} & \colhead{Close-in giant} & \colhead{$K>3\,\mps$} & \colhead{$K>1\,\mps$} \\
\colhead{} & \colhead{} & \colhead{intermediate} & \colhead{} & \colhead{} & \colhead{\& intermediate} & \colhead{\& $a<10\,$au} & \colhead{\& $a<1\,$au}
}
\startdata
    $N_0$ & 444 & 452 & 425 & 408 & 424 & 393 & 419\\
    $N_1$ & 28 & 17 & 41 & 47 & 42 & 56 & 43\\
    $N_2$ & 2 & 5 & 8 & 16 & 7 & 21 & 7 \\
    $N_3$ & 0 & 0 & 0 & 2 & 1 & 2 & 3\\
    $N_4$ & 0 & 0 & 0 & 1 & 0 & 1 & 2\\
    $N_5$ & 0 & 0 & 0 & 0 & 0 & 1 & 0\\
    $\bar{p}$ & 0.89 & 0.27 & 0.55 & 0.64 & 0.44 & 0.57 & 0.27\\
\enddata
\end{deluxetable*}

\begin{deluxetable*}{cccccccc}
\tablecaption{Intrinsic multiplicities for different classes of planets. Here values with uncertainties are the maximum-likelihood solution (Equation~\ref{eqn:solution_analytic}) and the associated 1-$\sigma$ error bars. 
For parameters that peak at zero, only the 2-$\sigma$ upper limits are provided. The last two rows are the estimated frequencies of planetary systems based on Equations~(\ref{eqn:cm-estimator}) and (\ref{eqn:fp-estimator}), respectively.
\label{tab:result}}
\tabletypesize{\small}
\tablehead{
\colhead{} & \colhead{\multirow{2}{*}{Close-in giant}} & \colhead{Close-in} & \colhead{\multirow{2}{*}{Cold giant}} & \colhead{\multirow{2}{*}{All giant}} & \colhead{Close-in giant} & \colhead{$K>3\,\mps$} & \colhead{$K>1\,\mps$} \\
\colhead{} & \colhead{} & \colhead{intermediate} & \colhead{} & \colhead{} & \colhead{\& intermediate} & \colhead{\& $a<10\,$au} & \colhead{\& $a<1\,$au}
}
\startdata
    $F_1$ & $6.7\pm1.2\%$ & $7.5\pm4.3\%$ & $12.6\pm2.5\%$ & $11.5\pm2.4\%$ & $14.9\pm3.6\%$ & $13.5\pm3.0\%$ & $22\pm8\%$ \\
    $F_2$ & $0.5^{+0.5}_{-0.3}\%$ & $7.5\pm3.1\%$ & $4.6\pm1.5\%$ & $6.5\pm1.8\%$ & $4.6\pm2.6\%$ & $10.2\pm2.4\%$ & $<17\%$ \\
    $F_3$ & \nodata & \nodata & \nodata & $<3.2\%$ & $1.3^{+1.7}_{-0.9}\%$ & $<3.8\%$ & $<13\%$ \\
    $F_4$ & \nodata & \nodata & \nodata & $0.7^{+0.9}_{-0.5}\%$ & \nodata & $<2.7\%$ & $9^{+3}_{-4}$ \\
    $F_5$ & \nodata & \nodata & \nodata & \nodata & \nodata & $1.0^{+0.9}_{-0.7}\%$ & \nodata \\
    \hline
    $F_{\rm p}$ & $7.2\pm1.6\%$ & $15\pm4\%$ & $17\pm3\%$ & $19.2\pm2.8\%$ & $21\pm4\%$ & $25\pm3\%$ & $33\pm7\%$ \\
    $\bar{n}_{\rm p}$ $^a$ & $0.076\pm0.013$ & $0.21\pm0.05$ & $0.22\pm0.03$ & $0.29\pm0.05$ & $0.28\pm0.05$ & $0.42\pm0.04$ & $0.58\pm0.09$ \\
    $\bar{m}_{\rm p}$ & $1.08\pm0.07$ & $1.5\pm0.3$ & $1.27\pm0.12$ & $1.52\pm0.15$ & $1.36\pm0.17$ & $1.63\pm0.16$ & $1.8\pm0.4$ \\
    \hline
    $F_{\rm p}^{\rm single}$ & $7.1\%$ & $17\%$ & $19\%$ & $22\%$ & $24\%$ & $30\%$ & $43\%$ \\
    $F_{\rm p}^{\rm multi}$ & $7.0\%$ & $12\%$ & $16\%$ & $18\%$ & $19\%$ & $23\%$ & $27\%$ \\
\enddata
\tablecomments{
$^a$ Here the values for the planet frequencies come from the derivation of Section~\ref{sec:planet-frequency}. The maximum-likelihood method yields consistent values, albeit with slightly different uncertainties. \\
}
\end{deluxetable*}

\begin{figure*}
    \centering
    \includegraphics[width=0.95\textwidth]{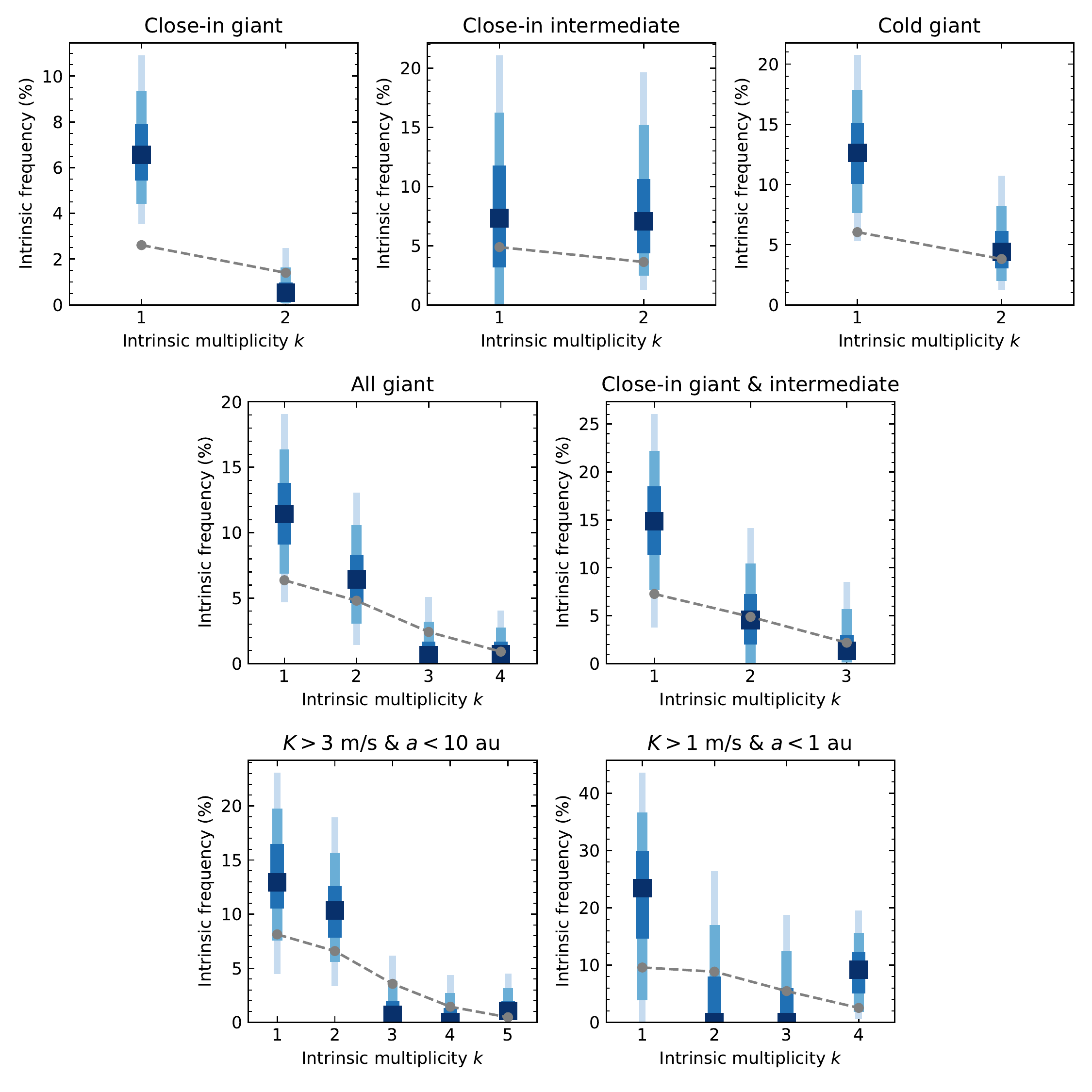}
    \caption{Illustrations of the derived intrinsic planet multiplicity distributions of the five chosen planet classes. The maximum-likelihood solutions and 1--3$\sigma$ confidence intervals are shown in blue colors with increasing transparency. For comparisons, we also show in each panel the scaled Poisson distribution to the total frequency of the given planetary system as the gray points. The shape parameter of the Poisson distribution is set to be the average multiplicity of the given planet type. Values for the maximum-likelihood solution and 1$\sigma$ confidence intervals (in some cases, the 2$\sigma$ upper limits) are tabulated in Table~\ref{tab:result}.}
    \label{fig:result}
\end{figure*}

We model the observed multiplicity distribution in the CLS Sun-like star sample following the method detailed in Section~\ref{sec:method}.
The results are summarized in Table~\ref{tab:result} and illustrated in Figure~\ref{fig:result}. Below we discuss the individual planet classes in details.

\subsubsection{Close-in giant planets}
With an average detection probability $\bar{p}=0.89$, the parameter space of the close-in giant planets is well covered by CLS. This suggests that the observed multiplicity distribution is close to the intrinsic multiplicity distribution. Indeed, our analysis reveals that $6.7 \pm 1.2\%$ and $0.5^{+0.5}_{-0.3}\%$ of Sun-like stars host one and two close-in giant planets, respectively. The intrinsic multiplicity fraction is very close to the observed multiplicity fraction.

Combining both multiplicities, we find that $7.2\pm1.6\%$ of Sun-like stars in the CLS sample host at least one giant planet within $1\,$au. The average multiplicity of such planets is marginally consistent with unity ($1.08\pm0.07$).

\subsubsection{Close-in intermediate-mass planets}

The CLS Sun-like star sample contains 17 and 5 systems with one and two close-in ($<1~$au), intermediate-mass ($10\,M_\oplus$--$0.3\,M_{\rm J}$) planets, respectively. Compared to the other planet classes, these planets are most susceptible to the sensitivity limit of CLS (Figure~\ref{fig:planets}), resulting in an average detection probability of $0.27$. The average number of close-in, intermediate-mass planets per Sun-like star is $0.21 \pm 0.05$. It is worth noting that the planet frequency rises rapidly as the planet mass goes below $\sim30\,M_\oplus$. As shown in Figure~\ref{fig:grid_map}, the integrated frequencies of close-in planets in the mass ranges of $30\,M_\oplus$--$0.3\,M_{\rm J}$ and 10--30$\,M_\oplus$ are $0.039 \pm 0.013$ and $0.16 \pm 0.04$, respectively.

As shown in Table~\ref{tab:result}, the maximum likelihood analysis reveals that $15 \pm 4\%$ of Sun-like stars in the CLS sample have close-in intermediate-mass planets, equally divided between one- and two-planet systems. The average multiplicity of such planets is $1.5 \pm 0.3$.

\subsubsection{Cold giants} \label{sec:cold-giants}

The CLS Sun-like sample has 41 and 8 systems with one and two cold giant planets, respectively, and the average detection probability is $0.55$. Here a cold giant planet is a planet with (minimum) mass above $0.3\,M_{\rm J}$ and semi-major axis in the range of 1--10\,au. According to this definition, our Saturn sits at (or very close to) the borderline, and thus the Solar System can be seen as hosting either one or two such cold giant planets. The frequency of cold giant planets around Sun-like stars is $0.22\pm0.03$, after the relevant grids in Figure~\ref{fig:grid_map} are combined.

As shown in Table~\ref{tab:result}, the maximum likelihood analysis reveals that $12.6 \pm 2.5\%$ and $4.6 \pm 1.5\%$ of Sun-like stars in the CLS sample have one and two cold giant planets, respectively. With both multiplicities combined, the total fraction of Sun-like stars have (at least one) cold giant planet is $17 \pm 3\%$, and each system has $1.27 \pm 0.12$ cold giant planets. These numbers suggest that about $27\%$ of systems with cold giant planets have two such planets and that, compared to the observed multiplicity rate ($8/49$), some $11\%$ of the CLS Sun-like stars with (so far) one known cold giant should have additional, currently undetected cold giant planet.

\subsubsection{All giant planets}

The CLS Sun-like sample contains 66 systems with at least one giant planet ($>0.3\,M_{\rm J}$)  within $10\,$au, including (16, 2, 1) hosting (2, 3, 4) such planets. Combining the average detection probabilities of close-in giant and cold giant planets, the average detection probability of a giant planet across the whole axis range ($\sim$0.03--10\,au) is $0.64$.

As shown in Table~\ref{tab:result}, the maximum likelihood analysis reveals that $19.2 \pm 2.8\%$ of Sun-like stars have at least one giant planet within 10\,au. This includes $11.5 \pm 2.4\%$ with only one giant planet, $6.5 \pm 1.8\%$ with two, $<3.2\%$ (2-$\sigma$ limit) with three, and $0.7^{+0.9}_{-0.5}\%$ with four. Taking all multiplicities together, each giant planet system hosts on average $1.52 \pm 0.15$ planets in the given parameter space. Compared to the observed multiplicity rate ($19/66=0.29$), the intrinsic multiplicity rate ($\sim0.4$) is substantially higher.

Out of the 19 systems with multiple giant planets, four contain giant planet pairs whose period ratios are close to ($<0.1$) first-order period commesurabilities: HD 128311, HD 37124, HD 82943, and HD 95128 (47 UMa).
This fraction, 4/19, is substantially higher than what one gets from a random sampling,
\footnote{For example, if the period ratio is randomly drawn from a log-flat distribution between 2 and 10, then one expects a $3\%$ chance to have a ratio between 2 and 2.1.}
Taking this fraction as the fraction of giant planet pairs in (or near) mean-motion resonances, we find that about 20\% of systems with multiple giant planets have apparent low-order period commensurabilities, which is in general consistent with previous studies \citep[e.g.,][]{Wright:2011}. However, it is noteworthy that the mean-motion resonance nature has not been confirmed in every of the four systems. Furthermore, giant planet pairs close to period commensurabilities are still rare, so our results based on the assumption that planet detections are independent from each other are largely unaffected.

\subsubsection{Close-in giant \& intermediate-mass planets}

There are (42, 7, 1) systems with (1, 2, 3) detected planets with semi-major axis $a<1\,$au and mass in the range of $10\,M_\oplus$--$13\,M_{\rm J}$ in the CLS Sun-like star sample. Combining the results of close-in giant and close-in intermediate-mass planets, the average detection probabilities of close-in planets in the joint parameter space is $0.44$.

As shown in Table~\ref{tab:result}, the maximum likelihood analysis shows that $21 \pm 4\%$ of Sun-like stars have close-in planets with masses above $10\,M_\oplus$. This includes $14.9 \pm 3.6\%$ of Sun-like stars hosting only one, $4.6 \pm 2.6\%$ hosting two, and $1.3^{+1.7}_{-0.9}\%$ hosting three such close-in planets.

\subsubsection{$K>3\,\mps$ planets within 10\,au}

The CLS Sun-like sample contains (56, 21, 2, 1, 1) stars with (1, 2, 3, 4, 5) planets whose orbits are inside $10\,$au and RV semi-amplitudes are above $3\,\mps$. The average detection probability, calculated in the $K$ vs.\ $a$ plane (see Figure~\ref{fig:grid_map_k}), is $0.57$.

As shown in Table~\ref{tab:result}, the maximum likelihood analysis reveals that $25 \pm 3\%$ of Sun-like stars in the CLS sample host at least one planet with $K>3\,\mps$ within $10\,$au. This includes $13.5 \pm 3.0\%$ with only one, $10.2 \pm 2.4\%$ with two, $<3.8\%$ with three, $<2.7\%$ with four, and $1.0^{+0.9}_{-0.7}\%$ with five such planets. Together with the measured frequency of such planets, $\bar{n}_{\rm p}=0.42\pm0.04$, each system hosts on average $1.63 \pm 0.16$ such planets. For comparison, the observed average multiplicity in the 81 CLS stars with such planets is 1.4. This suggests that about 19 planets with $K>3\,\mps$ and $a<10\,$au remain to be detected in this sample of planet hosts.

\subsubsection{$K>1\,\mps$ planets within 1\,au}

Within the CLS Sun-like sample, we find (43, 7, 3, 2) stars with (1, 2, 3, 4) planet detections whose orbits are inside $1\,$au and RV semi-amplitudes are above $1\,\mps$. The average detection probability is $0.27$, with the majority of the per-star detection probability below $0.2$.

As shown in Table~\ref{tab:result}, the maximum likelihood analysis shows that $33 \pm 7\%$ of Sun-like stars in the CLS sample host at least one planet with $K>1\,\mps$ and $a<1\,$au, with the majority ($22 \pm 8\%$) having only one such planet. The fraction of Sun-like stars with four such planets is found to be $9^{+3}_{-4}\%$. For frequencies of two- and three-planet systems, we find upper limits of $17\%$ and $13\%$, respectively. Together with the measured planet frequency of $0.58 \pm 0.09$, the average multiplicity of systems with such planets is determined to be $1.8 \pm 0.4$. Compared to the observed average multiplicity of $1.3$ in the 55 hosts of this type of planets, our results suggest that each system is currently missing on average $\sim$0.5 planets whose RV semi-amplitudes are above the current detection limit ($1\,\mps$). Future RV observations should be able to test this prediction.

\section{Discussion} \label{sec:discussion}

Below we discuss issues that may have impact on our derived results and compare our results with those from previous studies.

\subsection{Impact of the detection probability} \label{sec:p-impact}

\begin{figure}
    \centering
    \includegraphics[width=0.96\columnwidth]{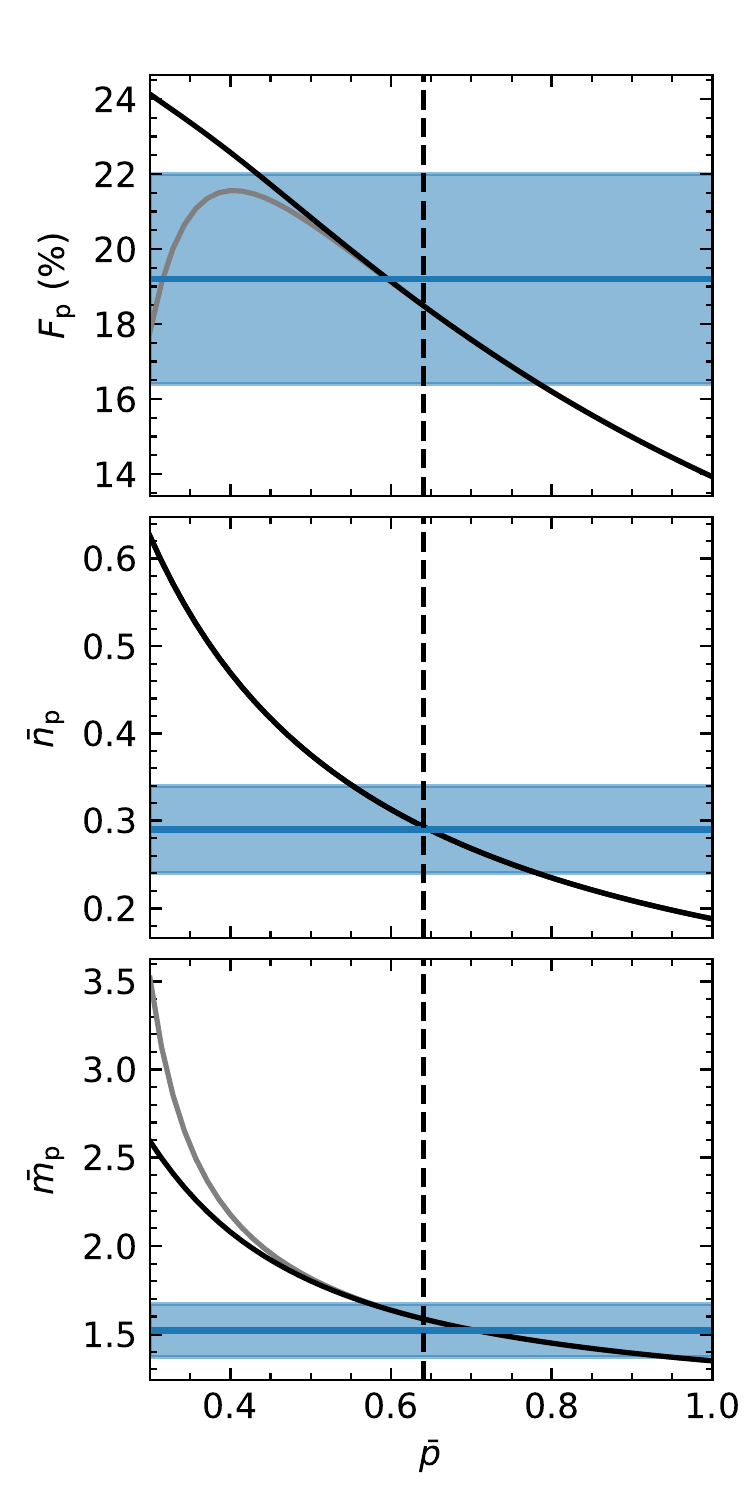}
    \caption{The fraction of Sun-like stars with giant planets (top panel), the average number of giant planets per Sun-like star (middle panel), and the average multiplicity of giant planets (bottom panel) as functions of the average detection probability $\bar{p}$. In each panel, the grey curve corresponds to the value from the analytic solution (Equation~\ref{eqn:solution_analytic}), the black curve is the maximum likelihood solution that takes into account the boundary conditions (i.e., $\bdv{F}>0$), and the horizontal line and shaded region are the derived frequency and the associated 1-$\sigma$ confidence interval, respectively. The vertical dashed line marks the value of $p$ that is adopted in the maximum likelihood analysis.}
    \label{fig:impact_of_p}
\end{figure}

The per-star detection probabilities, $\{p_l\}$, are crucial in the derivation of the intrinsic multiplicity distribution as well as the overall frequencies. As defined by Equation~(\ref{eqn:p_definition}), this per-star detection probability takes into account both the intrinsic planet distribution as well as the survey completeness. While the latter is (or can be) known fairly precisely, the former is usually not. As a result, the detection probability may not be so well determined.

Our derivation of the intrinsic multiplicity distribution makes use of the per-star detection probabilities, and thus it is less susceptible to the variation in the intrinsic planet distribution. Nevertheless, we investigate the impact of the mean detection probability parameter, $\bar{p}$, on the derived frequencies, as the intrinsic multiplicity distribution can be approximated as
\begin{equation} \label{eqn:solution_analytic}
    \bdv{F} \approx \frac{1}{N_\star} \mathcal{S}^{-1}(\bar{p}) \bdv{N} 
    = \frac{1}{N_\star} \mathcal{S}(\bar{p}^{-1}) \bdv{N} .
\end{equation}
The latter step makes use of the property of the sensitivity matrix \citep{Tremaine:2012}.  Figure~\ref{fig:impact_of_p} shows the derived frequencies---frequency of planetary systems, frequency of planets, and average multiplicity---as functions of the average detection probability $\bar{p}$ for the ``all giant planet'' case. We may convert the uncertainty in the frequency of planets, $\bar{n}_{\rm p}$, into the uncertainty of the average detection probability, given the direct connection between these two quantities (Equation~\ref{eqn:np}). This gives $\bar{p}_{\rm all~giants}=0.65\pm0.11$. The variation of $\bar{p}$ at such a level will not lead to substantial changes in the other two frequencies, both of which are rather insensitive to the variation of $\bar{p}$.

\subsection{Impact of maximum multiplicity} \label{sec:kmax}

\begin{figure*}
    \centering
    \includegraphics[width=0.9\textwidth]{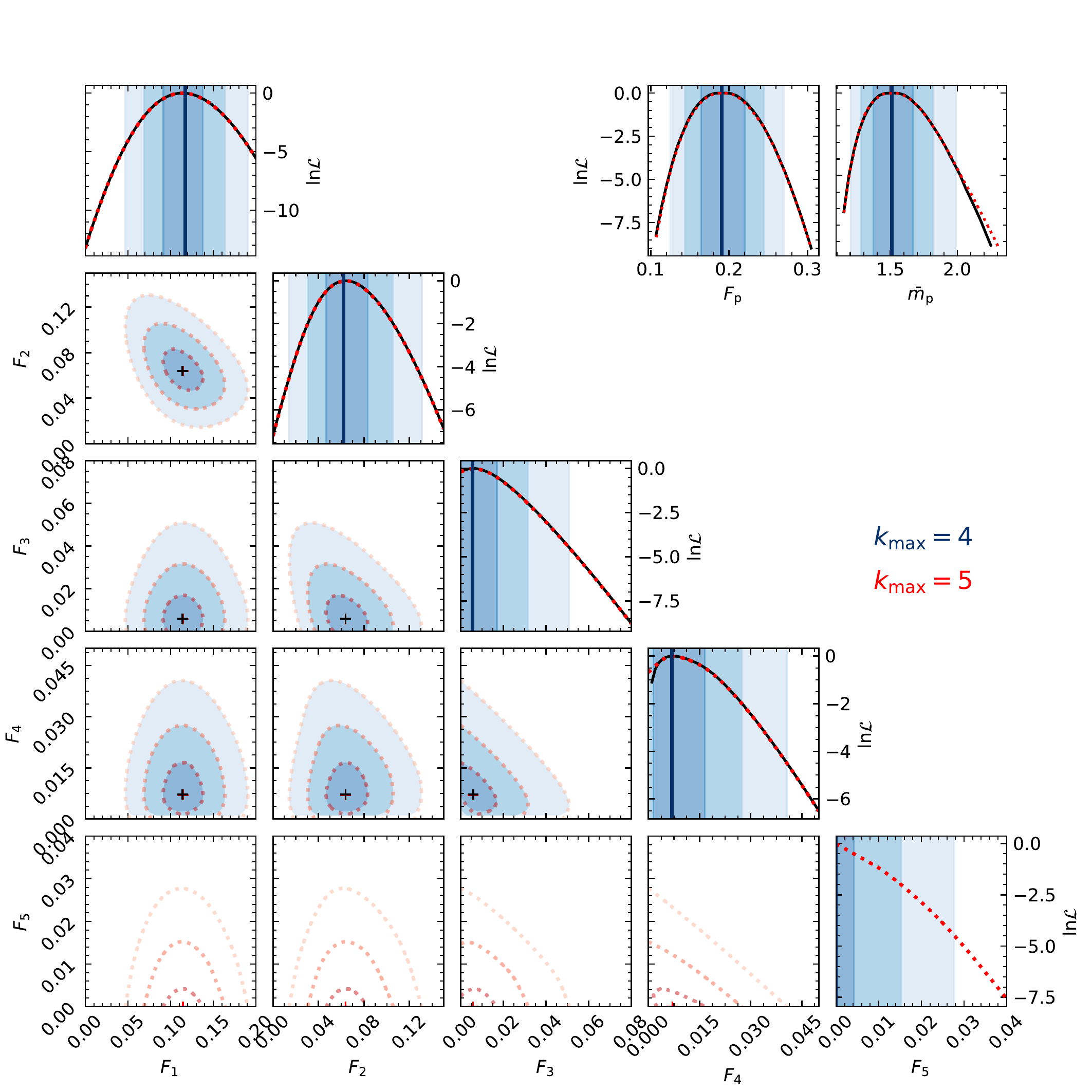}
    \caption{Likelihood contours between model parameters in the ``all giant planet'' case. We have used two different values for the maximum multiplicity, $k_{\rm max}$, in the modeling: Blue filled contours correspond to $k_{\rm max}=4$, which is the highest multiplicity with non-zero detections, whereas red dashed contours correspond to $k_{\rm max}=5$. For each set of contours, the plus signs indicate the analytic solution (Equation~\ref{eqn:solution_analytic}), and the contours with increasing levels of transparency correspond to the 1--3$\sigma$ uncertainty regions. The constraints on the frequency of planetary systems, $F_{\rm p}$, and the average multiplicity, $\bar{m}_{\rm p}$, are shown in the upper right corner. The inclusion of higher planet multiplicities with zero detections does not change the constraints on almost all low multiplicities. The only exception is seen in $F_4$, of which the lower limit is modified.}
    \label{fig:contours}
\end{figure*}

In our maximum-likelihood analysis, the multiplicity distribution is truncated at the highest multiplicity with non-zero detections. As explained in Section~\ref{sec:modeling}, this is because the inclusion of higher multiplicities in the data and the modeling introduces additional degrees of freedom but no benefit in the goodness of fit. 
Here we use the case of ``all giant planet'' to further demonstrate this point.

The CLS Sun-like sample has zero systems with more than four giant planets inside $10\,$au, and thus we have adopted the maximum multiplicity $k_{\rm max}=4$ in our maximum-likelihood analysis. The results have been reported in the previous section, and the likelihood contours are shown in Figure~\ref{fig:contours}. For comparisons, we also perform the analysis with $k_{\rm max}=5$. The resulting likelihood contours are also illustrated in Figure~\ref{fig:contours}. The inclusion of the higher multiplicity with zero planet detections does not change the maximum-likelihood solution or the uncertainties on all model parameters except for $F_4$, of which the lower limits are slightly modified. This is because the introduction of $F_5$ permits $F_4=0$, which is otherwise not allowed with $k_{\rm max}=4$. The integrated fraction of planetary systems and the average multiplicity remain largely unaffected, too.

\subsection{Frequencies of giant planets}

Regarding the frequencies of giant planets around Sun-like stars, we find
\begin{equation}
\left\{
\begin{array}{ll}
    \bar{n}_{\rm p} ({\rm HJ}) & = 0.028 \pm 0.008 \\
    \bar{n}_{\rm p} ({\rm WJ}) & = 0.048 \pm 0.011 \\
    \bar{n}_{\rm p} ({\rm CJ}) & = 0.22 \pm 0.03 \\
\end{array} \right. .
\end{equation}
Here ``HJ'', ``WJ'', and ``CJ'' denote hot Jupiters (giant planets within 0.1\,au), warm Jupiters (giant planets between 0.1--1\,au), and cold Jupiters (giant planets between 1--10\,au), respectively. The HJ rate we derive is higher than previous RV studies, which collectively points to a value of $\sim1\%$ \citep[e.g.,][]{Cumming:2008, Mayor:2011, Wittenmyer:2020}. However, given the large statistical uncertainties and the possibly different stellar distributions (e.g., metallicity, binarity) in these studies, we do not investigate further the cause of this discrepancy. 

Regarding the WJ and CJ rates, a very direct comparison is with \citet{Fulton:2021}, who used the whole CLS sample and derived the frequency of planets around FGKM stars.
\footnote{Note the different definition of a ``giant planet'' used in \citet{Fulton:2021} and here. \citet{Fulton:2021} set the lower limit of a ``giant planet'' to be $30\,M_\oplus$.}
Adopting their best-fit broken power-law model, we find $\sim0.05$ and $\sim0.14$ for the frequencies of WJs and CJs, respectively. While the WJ rate is consistent with ours, their CJ rate is substantially lower. This is probably because \citet{Fulton:2021} included M-type stars in their analysis. In the CLS sample, there are only 13 cold giant planets found around 245 M-type stars, whereas the numbers are 57 and 474 for Sun-like stars. Our rates of WJs and CJs are also in general agreement with studies using different samples \citep[e.g.,][]{Mayor:2011, Fernandes:2019, Wittenmyer:2020}. For example, the re-analysis of the \citet{Mayor:2011} sample by \citet{Fernandes:2019} found $\sim0.04$ and $\sim0.17$ for frequencies of WJs and CJs, respectively. These values are estimated based on the asymmetric broken power-law model used in \citet{Fernandes:2019}.

The relative frequencies between HJ and WJ can provide useful insights into the formation of these close-in giant planets (see \citealt{Dawson:2018} and references therein). In the CLS Sun-like star sample, the overall rate of WJ is higher than the overall rate of HJ, and the rate of WJ per log interval is within a factor of two of that of HJ. Both features pose challenges to the theoretical models of HJ formation.

\subsection{Frequencies of planetary systems}

One approach to derive the frequency of planetary systems in RV surveys, which has been frequently used in the literature, is to include only one out of the multiple planet detections in each system and regard the resulting frequency of planets as the frequency of planetary systems \citep[e.g.,][]{Cumming:2008, Mayor:2011}. In the mathematical framework of the present work, this is
\begin{equation} \label{eqn:cm-estimator}
    F_{\rm p}^{\rm single} = \frac{1}{N_\star \tilde{p}} \sum_{j=1} N_j .
\end{equation}
Here $\tilde{p}$ is the average detection probability derived from the planet sample that includes at most one planet detection for each star. It will not be too different from the value of $\bar{p}$, given the dominating number of single-detections in typical RV samples.

For our approach, which takes the planet multiplicity fully into account, a reasonable approximation of the frequency of planetary systems is
\begin{equation} \label{eqn:fp-estimator}
    F_{\rm p}^{\rm multi} = \frac{1}{N_\star p_m} \sum_{j=1} N_j .
\end{equation}
Here the correction factor, $p_m$, quantifies the probability of detecting at least one planet in a system with $\bar{m}_{\rm p}$ planets
\begin{equation}
    p_m = 1-(1-\bar{p})^{\bar{m}_{\rm p}} .
\end{equation}
The estimator given by Equation~(\ref{eqn:cm-estimator}) yields unbiased rates only when the average multiplicity $\bar{m}_{\rm p}$ approaches unity. We provide in the last two rows of Table~\ref{tab:result} the two frequency estimators for our chosen planet types. In all cases, our new estimator, $F_{\rm p}^{\rm multi}$, yields better agreement with the true frequency. 

The estimator given by Equation~(\ref{eqn:fp-estimator}) has broader implications to other statistical analyses that involve planet multiplicity issues. For Sun-like stars, the sensitivity region of the \emph{Kepler} transit mission is almost fully enclosed in the box-shaped parameter space defined by a lower radius limit of $\sim1\,R_\oplus$ and an outer period limit of $\sim1\,$yr. According to the estimated planet frequencies in this box-shaped region \citep[e.g.,][]{ZhuDong:2021}, the probability of having any one planet inside the \emph{Kepler} sensitivity region is $\bar{p}_{\rm kep}\approx0.8$. For \emph{Kepler}-like planetary systems with an average multiplicity of three \citep{Zhu:2018}, the chance to see at least one planet out of the box-shaped parameter space to be inside the \emph{Kepler} sensitivity region is therefore $1-(1-\bar{p}_{\rm kep})^3=0.99$. In other words, the fraction of Sun-like stars with planets of radii above $R_\oplus$ and periods within 1 yr is almost the same as the fraction of Sun-like stars with at least one planet inside the \emph{Kepler} sensitivity range. For comparison, the sensitivity correction on the planet-by-planet basis would lead to an increase in the frequency of planetary systems by nearly a factor of two \citep[e.g.,][]{Yang:2020}.

In addition to providing an unbiased estimate of the total frequency of planetary systems, the approach undertaken in the present work also yields the intrinsic multiplicity distribution $\bdv{F}$, which is more informative to theoretical models as well as other statistical studies. For example, for giant planets within $3\,$au, based on which \citet{Cumming:2008} derived a total frequency of $10.5\%$, our approach yields the following total rates based on the CLS Sun-like sample
\begin{equation}
\left\{
\begin{array}{ll}
    F_{\rm p}^{\rm C08} & = 12.2 \pm 2.1\% \\
    \bar{n}_{\rm p}^{\rm C08} & = 0.15 \pm 0.03 \\
    \bar{m}_{\rm p}^{\rm C08} & = 1.21 \pm 0.10 \\
\end{array} \right. .
\end{equation}
as well as the intrinsic multiplicity distribution
\begin{equation}
    F_{k=1,2,3}^{\rm C08}= (10.6 \pm 1.5\%,~ 0.8^{+0.6}_{-0.4}\%,~ 0.9^{+0.5}_{-0.4}\%) .
\end{equation}
The overall rate ($F_{\rm p}^{\rm C08}$) is higher than \citet{Cumming:2008} probably because of the difference in the stellar samples.

\subsection{Correlations (or not) between different planet populations}

By including in the analysis the intrinsic multiplicity distribution, our method does not require the presences of multiple planets in the same system to be independent (although the properties of them should be, largely, independent). Therefore, it allows us to investigate whether or not the existence of a certain type of planet depends on the existence of some other type of planet in the same system. Considering the relatively large uncertainties of the derived frequencies, here we focus on confirming previously found (anti-)correlations rather than reporting tentative evidence for new ones.

Under the assumption that the presences of close-in and cold giant planets are not correlated, the frequency of planetary systems with any giant planets within 10\,au should be $1-(1-7.2\%) \times (1-17\%)= 23\%$. This is higher than what is given by the maximum likelihood analysis over the joint parameter space ($19.2\%$, see Table~\ref{tab:result}). Therefore, the assumption is rejected/disfavored and the close-in and cold giant planets are positively correlated. This is consistent with previous findings that close-in giant planets frequently have distant companions \citep[e.g.,][]{Knutson:2014, Bryan:2016}. Similarly, our results also suggest positive correlations between close-in planets (close-in giant and intermediate-mass planets) and cold giant planets, which also confirms previous findings \citep[e.g.,][]{ZhuWu:2018, Bryan:2019, Rosenthal:2022}.

It is noticed that the fraction of stars with close-in giant and intermediate-mass planets equals the summation of frequencies of ``close-in giant'' and ``close-in intermediate-mass'' planets. Indeed, under the assumption that the presences of these latter two classes of planets are uncorrelated, one would expect that the frequency of close-in CLS planetary system should be $1-(1-0.072) \times (1-0.15)=21\%$, which is consistent with the fraction of planets in the joint parameter space ($21\pm4\%$). In other words, the two planet populations---close-in giant planets and close-in intermediate-mass planets---are uncorrelated. This may be explained by the combination of two different features of close-in planets. On the one hand, the majority of HJs and some WJs are lonely \citep[e.g.,][]{Steffen:2012, Hord:2021}. On the other hand, at least half of WJs do have nearby, small companions \citep{Huang:2016}.

\section{Summary}

Our present study develops a general method that can recover the intrinsic multiplicity distribution of planets out of observational surveys with incomplete coverage of the planetary parameter space. We then applied it to the CLS Sun-like sample. The key results are summarized in Table~\ref{tab:result} and illustrated in Figure~\ref{fig:result}. A few notable findings to highlight here:
\begin{itemize}
    \item About $19.2\%$ of Sun-like stars in the CLS sample host giant ($>0.3\,M_{\rm J}$) planets within 10\,au, with the majority having such planets in the cold ($>1\,$au) region. Of the giant planet hosts, about 40\% host at least two such planets.
    \item Within the 1\,au region, giant planets typically do not find siblings of the same kind, whereas lower-mass ($10\,M_\oplus \mbox{--} 0.3\,M_{\rm J}$) planets usually do. In total, $21\%$ of Sun-like stars host planets with masses above $10\,M_\oplus$ within 1\,au.
    \item About $25\%$ of Sun-like stars host planets with semi-major axis $a<10\,$au and RV semi-amplitude $K>3\,\mps$, and about half of them have at least two such planets. With a RV precision of $1\,\mps$, about 0.5 planets remain to be detected around each of the CLS Sun-like hosts.
    \item The CLS Sun-like sample reports a hot Jupiter rate of $2.8 \%$, which is a factor of $\sim$3 higher than what previous studies have found from different star samples.
\end{itemize}

It is worth noting that, although our method is developed for the RV method, it is generally applicable to other exoplanet-finding techniques such as gravitational microlensing, astrometry, and direct imaging. It also works for the transit method after the geometric transit probability is properly taken into account \citep{Tremaine:2012}.

\begin{acknowledgments}
I would like to thank Fei Dai and Scott Tremaine for comments and suggestions on an earlier version of the manuscript. This work is supported by the National Science Foundation of China (grant No.\ 12173021 and 12133005).
\end{acknowledgments}

\bibliography{my_bib}{}
\bibliographystyle{aasjournal}



\end{CJK*}
\end{document}